  \documentstyle[prd, preprint,eqsecnum,aps]{revtex}  
  \begin{document}
  \draft
   \preprint{\vbox{
  \hbox{CTP-TAMU-17/99, SINP-TNP/99-16, hep-th/9905056}
  }}

  \title{(F, D5) Bound State, SL(2, Z) Invariance and The Descendant\\
         States in Type IIB/A String Theory\\}

  \author {J. X. Lu$^1$ and Shibaji Roy$^2$}
  \address{$^1$Center for Theoretical Physics,
  Texas A\&M University, College Station, TX 77843\\
  $^2$Saha Institute of Nuclear Physics,
  1/AF Bidhannagar, Calcutta 700 064, India\\
E-mail: jxlu@rainbow.physics.tamu.edu, roy@tnp.saha.ernet.in}

  \maketitle
  
  \begin{abstract}
  
Recently the space-time configurations of a set of non-threshold
bound states, called the (F, Dp) bound states, have been constructed
explicitly for every $p$ with $2 \le p \le 7$ in both type IIA (for $p$ even)
and type IIB (for $p$ odd) string theories by the present authors. By making
use of the SL(2, Z) symmetry of type IIB theory we construct a more 
general SL(2, Z) invariant bound state of the type ((F, D1), (NS5, D5))
in this theory from the (F, D5) bound state. There are actually an infinite
number of $(m,n)$ strings forming bound states with  $(m',n')$
5-branes, where strings lie along one of the spatial directions of the
5-branes. By applying T-duality along one of the transverse 
directions we also construct the bound state ((F, D2), (KK, D6)) 
in type IIA string theory. Then we give a list of possible bound states 
which can be obtained from these newly constructed bound states by applying
T-dualities along the longitudinal directions as well as S-dualities to those 
in type IIB theory.
\end{abstract}
 \newpage

  \section{Introduction\protect\\}
  \label{sec:intro}
This is in sequel to our series of study on a new kind of bound states
that exist in both type IIA and type IIB string theories. In 
\cite{lurone}, we have
provided arguments from the worldvolume point of view, that there exist
BPS bound states of Dp branes carrying certain units of quantized constant
electric fields, called the (F, Dp) bound states, for every $p$ with
$1 \le p \le 8$ in type IIA (for $p$ even) and type IIB (for $p$ odd)
string theory. The space-time configurations of these bound states
have been constructed explicitly for $2 \le p \le 7$ in 
ref.[2]\footnote{The
configurations for $p = 3, 4, 6$ were also given previously in 
\cite{rust,grelpt,cosp}, respectively. Similar non-threshold bound states
in M theory or type IIA/IIB theory for a $p'$-brane within another $p$-brane
with $p'< p$ were studied in \cite{rust,grelpt,cosp,schone,polr}.}. Each of 
these bound states preserve one half of the space-time supersymmetries. In the 
worldvolume picture the F in (F, Dp) represents the uniform and constant
electric field lines flowing along, say, $x^1$ axis of the Dp-brane due to the
uniform and homogeneous charge distribution on the rest of the 
$(p-1)$ plane placed
at $x^1 = -\infty$, originating from an infinite number of open strings
ending on this surface. On the other hand, the space-time configuration
allows us to identify these field lines with the infinitely long fundamental
strings or F-strings in the bulk. To make this identification more
concrete, we have calculated the charges carried by F-strings, Dp-branes
as well as the mass per unit $p$-brane volume and have shown that they 
match precisely with what we expect from the worldvolume study. We have
noted in [1], that since type IIB theory is conjectured to possess an 
SL(2, Z) invariance, there must exist more general bound states than (F, Dp)
in this theory. By making use of this observation, we have constructed
the more general non-threshold bound state of the type ((F, D1), D3) 
and some of its
T-dual descendants in ref.[8].

In this paper, we make an SL(2, Z) transformation on the non-threshold
bound state (F, D5) in type IIB theory to construct ((F, D1), (NS5, D5))
bound state. The space-time configuration consisting of the metric, the
dilaton, the axion and the other non-vanishing gauge fields for this 
bound state are constructed explicitly. The initial (F, D5) 
configuration consists
of an infinite number of NS strings (each NS string is actually $q$ F-strings) 
distributed uniformly over $s$ D5-branes and lying along
one of the spatial directions of D5-brane, where $q$ and $s$ are relatively
prime integers as discussed in \cite{lurtwo}. We here consider a genuine 
initial (F, D5) bound states, i.e., both $q$ and $s$ are 
non-zero.  
In general, we expect that in the
bound state ((F, D1), (NS5, D5)), there are infinite number of $(m,n)$
strings lying along one of the spatial directions of $(m',n')$ 5-branes.
Although for the degenerate case when either the strings or the
5-branes (but not both) are present the integers $(m, n)$ and $(m', n')$ 
are individually
relatively prime, for the general non-degenerate case this is not necessarily
so. For the general ((F, D1), (NS5, D5)) bound state,  
i.e. when 
the integers $m, n$ and  $m', n'$ are non-zero,  we find that 
a consistent quantization of
the charges associated with the NSNS and RR gauge fields of
both the strings and the 5-branes relates the charges of the strings 
with those of the 5-branes. As a result the integers
$(m,n)$ corresponding to the electric charges 
of the strings and the integers
$(m',n')$ corresponding to the magnetic charges of 
the 5-branes get related as $(m ,n) = k(a, b)$, $(m', n') = k'(-b, a)$,
where $(a, b)$ and $(k, k')$ are relatively prime integers. This
fact in turn tells us that when both $k$, $k'$ are non-zero the existence 
of bound states between $m$
fundamental strings and $n$ D-strings may imply the existence of bound states
between $m'$ NS5-branes 
and $n'$ D5-branes, where the integers $(m, n)$ and $(m', n')$ are related
to each other as given before. Thus we find that in general the SL(2, Z) 
invariant bound state 
((F, D1), (NS5, D5)) of type IIB theory is characterized by 
two pairs of relatively
prime integers $(a, b)$ and $(k, k')$. We can obtain the other bound states,
namely, (F, D5) and (D1, NS5) from this general solution by setting $a=1$,
$b=0$ and $a=0$, $b=1$ respectively. Also note that the degenerate
(NS5, D5) and (F, D1) cases can be obtained from the general 
((F, D1), (NS5, D5)) bound state by setting (i) $k=0$, $k'=1$ and (ii)
$k'=0$, $k=1$. For the former case we get the SL(2, Z) 5-branes discussed
in [9], whereas for the latter case we get SL(2, Z) strings obtained in
[6] with four additional isometries. 
But because the charges of the strings and the 5-branes are
related as mentioned above, we cannot have bound states of the form (F, NS5)
and (D1, D5)  consistent with the fact that these bound states preserve
1/4 rather than 1/2 of the spacetime supersymmetries. We have also obtained
the expression for the tension of SL(2, Z) invariant non-threshold bound
state ((F, D1), (NS5, D5)) and have shown how it reduces to the tensions for
the corresponding special case bound states.

The descendants of this bound state could be obtained by
applying  T-duality along various transverse and longitudinal
directions.  We give an explicit construction of the
bound states ((F, D2), (KK, D6)) in type IIA theory 
by applying T-duality in one of the transverse 
directions. We also discuss how the bound states (F, D6) and (D2, KK) as well
as the degenerate cases (F, D2) and (KK, D6) can be obtained from this general
bound state as special cases. The tension expression for the bound
state ((F, D2), (KK, D6)) is also given.
We point out
the problem of taking further T-dualities along the transverse directions.
Finally, we give a list of
possible other descendant bound states which can be obtained from these by
T-dualities in various longitudinal
directions as well as S-dualities to those in type IIB theory.

This paper is organized as follows. In section 2, we use the SL(2, Z)
invariance of type IIB theory to construct the non-threshold
((F, D1), (NS5, D5)) bound state starting from the (F, D5) one. We show
that (F, D5), (D1, NS5), (F, D1) and (NS5, D5) bound states can be obtained
from this bound state as special cases. In section 3, we apply T-duality
on ((F, D1), (NS5, D5)) bound state
along one of the transverse directions to construct 
((F, D2), (KK, D6))
bound state and discuss the special cases as in section 2. 
Since we have shown how to implement
S- and T-dualities in general, we here list all the possible bound states
that can be obtained from the abovementioned bound states 
by the application of
T-dualities in various longitudinal directions and S-dualities to those
belonging to type IIB theory.
We conclude this paper in section 4.

\newpage
  
\section{SL(2, Z) Invariance and Non-Threshold ((F, D1), (NS5, D5))
Bound State\protect\\}
  \label{sec:RDP} 
  
In this section, we will use the SL(2, Z) symmetry of type IIB theory to
construct the non-threshold bound state ((F, D1), (NS5, D5)) from the
explicit solution (F, D5) given in \cite{lurtwo}. We will follow the
procedure outlined in \cite{schone}, \cite{rone},\cite{lur}. Let us begin 
with (F, D5)
solution \cite{lurtwo} given by the metric,
\begin{eqnarray}
d s^2 =&& H'^{1/2} \, H^{1/4} \,
\left[ H^{-1} \left(- (d x^0)^2 +
(d x^1)^2\right)
+ H'^{-1} \left( (d x^2)^2 + (d x^3)^2 
+ (d x^4)^2 + (d x^5)^2\right)\right.\nonumber\\
&&\qquad\qquad \left. + d y^i d y^i\right],
\label{eq:iibmfive}
\end{eqnarray}
with $i = 1, 2, 3, 4$; the dilaton,
\begin{equation}
e^{\phi} = H^{- 1/2},
\label{eq:iibdfive}
\end{equation}
and the remaining non-vanishing fields,
\begin{eqnarray}
H_3^{(1)} &=&  - q \Delta_{(q,s)}^{-1/2}\,  
d H^{-1}\wedge d x^0 
\wedge d x^1, \nonumber\\     
H_3^{(2)} &=& s \,\frac{\sqrt{2} \kappa_0 Q_0^5}{\Omega_3}\, 
\epsilon_3,\nonumber\\
H_5 &=&  qs  \Delta_{(q,s)}^{-1} 
\, H'^{-2} \,d H \wedge d x^2 \wedge d x^3 \wedge d x^4 
\wedge d x^5.
\label{eq:iibrffive}
\end{eqnarray}
Note that in writing this solution we have set the scalars $\phi_{B0} =
\chi_{B0} = 0$, as they have nothing to do with the dilaton and the
axion in the theory [2].  Also in the above $H_3^{(1)}$ and
$H_3^{(2)}$ are the NSNS and RR 3-form field strengths. $H_5$ is the
self-dual 5-form field-strength with $H_5 = \ast H_5$, where $\ast$
denotes the Hodge dual. $H$ is a harmonic function given by
\begin{equation}
H = 1 + \frac{Q_5}{r^2},
\end{equation}
where $r^2 = y^i y^i$ and $Q_5 = \Delta_{(q,s)}^{1/2} \sqrt{2} \kappa_0
Q_0^5/(2\Omega_3)$, $H'$ is another harmonic function defined as
\begin{equation}
H' = \frac{q^2 + s^2 H}{\Delta_{(q,s)}} = 1 + \frac{s^2 Q_5 
/ \Delta_{(q,s)}}{r^2},
\end{equation}
with $\Delta_{(q,s)} = q^2 + s^2$. Here $q$ and $s$ are two relatively 
prime integers denoting respectively the quantized NS string charge or
the number of F-strings per $(2\pi)^4 \alpha'^2$ of four dimensional area
perpendicular to the strings in (F, D5) and the D5-brane charge as 
discussed in \cite{lurtwo}. $\epsilon_n$ denotes the volume form on an
$n$-sphere whereas $\Omega_n$ is the volume of a unit $n$-sphere and is 
given as,
\begin{equation}
\Omega_n = \frac{(2\pi)^{(n+1)/2}}{\Gamma(\frac{n+1}{2})}.
\end{equation}
Also, $\sqrt{2}\kappa_0 = (2\pi)^{7/2} \alpha'^2$ and the unit charge for a 
Dp-brane is $Q_0^p \equiv (2\pi)^{(7-2p)/2} \alpha'^{(3-p)/2}$.

The electric charge of the F-strings in (F, D5) bound state can be calculated
as,
\begin{equation}
e^{(1)} = \frac{1}{\sqrt{2}\kappa_0} \int_{R^4 \times S^3} \left(
e^{-\phi} \ast H_3^{(1)} + H_3^{(2)} \wedge B_4\right).
\end{equation}
But as mentioned in \cite{lurtwo}, this expression is in fact infinite
as there are infinite number of F-strings in (F, D5). However, we can
still define a quantized charge from (2.7) in the form as given below
\cite{lurtwo}
\begin{equation}
Q^{(1)} = (2\pi)^4 \alpha'^2 \frac{e^{(1)}}{\sqrt{2}\kappa_0 A_4} =
 q T_f,
\end{equation}
with $T_f = 1/(2\pi \alpha')$ the fundamental string tension.
Here $A_4 = \int dx^2 dx^3 dx^4 dx^5$ is the coordinate area of 
$x^2x^3x^4x^5$-plane. The charge $Q^{(1)}$ represents 
the number of F-strings per $(2\pi)^4 \alpha'^2$ area over 
$x^2x^3x^4x^5$-plane measured in some units (Note the F-strings lie 
along the $x^1$-axis). 
Also the quantized magnetic charge of the D5-brane is given as,
\begin{equation}
P^{(2)} = g^{(2)} = \frac{1}{\sqrt{2}\kappa_0} \int_{S^3} H_3^{(2)} =
s Q_0^5
\end{equation}
It is well-known that type IIB supergravity possesses a classical
Cremmer-Julia \cite{CJ} symmetry group SL(2, R). A discrete 
subgroup
SL(2, Z) is now believed \cite{HT} to survive in the full quantum
type IIB string theory. Under a global SL(2, R) symmetry the Einstein
metric $g_{\mu\nu}$ is a singlet, the two 3-form field strengths $H_3^{(1)}$
and $H_3^{(2)}$ transform as a doublet and the 5-form field strength is
also a singlet. So, the transformations of the various fields alongwith
the two scalars, the dilaton ($\phi$) and the axion ($\chi$, the RR scalar)
parametrizing the coset SL(2, R)/SO(2) defined as ${\cal M} =
e^{\phi} \left(\begin{array}{cc}\chi^2 + e^{-2\phi} 
&\qquad \chi\\ \chi &\qquad 1
\end{array}\right)$ are:
\begin{eqnarray}
g_{\mu\nu} &\rightarrow & g_{\mu\nu}, \quad {\cal M} \rightarrow \Lambda
{\cal M} \Lambda^T, \quad \left(\begin{array}{c} H_3^{(1)}\\H_3^{(2)}
\end{array}\right) \equiv {\cal H} \rightarrow (\Lambda^{-1})^T 
{\cal H}\nonumber\\
H_5 &\rightarrow & H_5
\end{eqnarray}
where $\Lambda$ is a global SL(2, R) transformation matrix and `$T$' denotes
the transpose of a matrix.

Let us next look at how the charges would transform under the global SL(2, R)
transformation. Since a general SL(2, Z) invariant configuration will have
both (F, D1) strings (infinite numbers of them) living on (NS5, D5) branes,
the charge expression in (2.7) will be modified to give electric charges of
both F-string and D-string as,
\begin{equation}
e^{(i)} = \frac{1}{\sqrt{2}\kappa_0} \int_{R^4 \times S^3} \left(
{\cal M}^{ij} \ast H_3^{(j)} + \epsilon^{ij} H_3^{(j)} \wedge B_4\right)
\end{equation}
where $i, j = 1, 2$ and $\epsilon^{ij}$ is the SL(2, R) invariant totally
antisymmetric tensor with $\epsilon^{12} = 1$. As before, $e^{(i)}$ is
not well-defined and we can define the quantized charges as,
\begin{equation}
Q^{(i)} = \frac{(2\pi)^4 \alpha'^2 e^{(i)}}{\sqrt{2}\kappa_0 A_4} 
\end{equation}
The quantized magnetic charges of the NS5-brane and D5-brane can be obtained
as,
\begin{equation}
P^{(i)} = g^{(i)} = \frac{1}{\sqrt{2}\kappa_0} \int_{S^3} H_3^{(i)} 
\end{equation}
Note that the electric charge in (2.11) or (2.12) is a Noether charge
and follows from the equation of motion whereas the magnetic charge is
topological and follows from Bianchi identity. It is clear from (2.10)
that the electric charges of (F, D1) strings and the magnetic charges of
(NS5, D5) branes would transform as,
\begin{equation}
\left(\begin{array}{c}Q^{(1)}\\Q^{(2)}\end{array}\right) \equiv {\cal Q}
\rightarrow \Lambda {\cal Q};\qquad 
\left(\begin{array}{c} P^{(1)}\\P^{(2)}\end{array}\right) \equiv {\cal P}
\rightarrow (\Lambda^{-1})^T {\cal P} 
\end{equation}
Now in order to obtain the global SL(2, R) transformation matrix $\Lambda_0$,
we start with the zero asymptotic values of the dilaton and the axion i.e.
${\cal M}_0^{(initial)} = I$, where $I$ is the identity matrix and demand
that $\Lambda_0$ will transform it into a fixed but arbitrary value as
\begin{equation}
{\cal M}_0 = \Lambda_0 I \Lambda_0^T
\end{equation}
where
${\cal M}_0 =
e^{\phi_0} \left(\begin{array}{cc}\chi_0^2 + e^{-2\phi_0} &\qquad 
\chi_0\\ \chi_0 &\qquad 1
\end{array}\right)$, with $\phi_0$ and $\chi_0$ denoting the 
arbitrary but given
asymptotic values of the scalars. Eq.(2.15) will fix the SL(2, R) matrix
$\Lambda_0$ in terms of $\phi_0$, and $\chi_0$ and an undetermined SO(2) 
angle $\alpha$ as,
\begin{equation}
\Lambda_0 = e^{\phi_0 /2}\left(\begin{array}{cc}
             e^{-\phi_0}\, \cos\,\alpha + \chi_0\,\sin\,\alpha 
\qquad& - e^{-\phi_0}\,\sin\,\alpha \\
	     \sin\,\alpha \qquad& \cos\,\alpha \end{array}\right).
\label{eq:sl2rm0}
\end{equation}
The angle $\alpha$ will be given shortly.

Note that once we apply SL(2, R) transformation on the initial quantized
charges (2.8) and (2.9) of (F, D5) by Eq.(2.14), the charges will no
longer remain quantized. In order to get around this problem, one either
needs to introduce compensating factors in place of both the charges
$q$ and $s$ or replace $q$ and $s$ by arbitrary classical charges as
${\tilde \Delta}^{1/2}_{(m,n)}$ and ${\bar \Delta}^{1/2}_{(m',n')}$
respectively, where $\Delta$'s are the arbitrary numbers and will be
determined in the process of charge quantization \cite{NT}. 
By imposing that the
transformed charges are integers\footnote{When we consider either $(m, n)$
 strings \cite{schtwo} or $(m', n')$ 5-branes \cite{lurf}, 
 the integers $(m,n)$ or $(m',n')$ need to be
 relatively prime in order for the
strings  or 5-branes to form non-threshold bound
states. For the present non-threshold bound state consisting of an infinite
number of $(m,n)$ strings and a $(m', n')$ 5-brane,
  $(m, n)$ and
$(m', n')$ are not necessarily relatively prime as we will see later.}, 
namely, $(m,n)$ for strings and $(m',n')$ for the 
5-branes we have from (2.14)
\begin{equation}
 \left(\begin{array}{c}
      m\\
      n \end{array} \right) = \Lambda_0^1   \left(\begin{array}{c}
     \tilde{\Delta}_{(m,n)}^{1/2} \\
      0 \end{array} \right).
\end{equation}
for the strings and
\begin{equation}
 \left(\begin{array}{c}
      m'\\
      n' \end{array} \right) = ((\Lambda_0^5)^{-1})^T   
\left(\begin{array}{c} 0 \\
     \bar{\Delta}_{(m',n')}^{1/2} 
       \end{array} \right).
\end{equation}
for the 5-branes. 
$\Lambda_0^1$ and
$\Lambda_0^5$ are the transformation matrices for strings and 5-branes.
Eqs.(2.17) and 
(2.18) determine
the form of $\Lambda_0^1$ and $\Lambda_0^5$ in terms of the asymptotic values
of the dilaton $\phi_0$ and the axion $\chi_0$ as follows:
\begin{equation}
\Lambda_0^1 = \frac{1}{\tilde{\Delta}_{(m,n)}^{1/2}} \left(\begin{array}{cc}
m \qquad& -  n \,e^{-\phi_0} + \chi_0 \,(m - \chi_0 n) \,e^{\phi_0} \\
n\qquad & (m - \chi_0 n)\, e^{\phi_0} \end{array} \right),
\end{equation}
and
\begin{equation}
\Lambda_0^5 = \frac{1}{\bar{\Delta}_{(m',n')}^{1/2}} \left(\begin{array}{cc}
n' \qquad&   m' \,e^{-\phi_0} + \chi_0 \,(n' + \chi_0 m') \,e^{\phi_0} \\
-m'\qquad & (n' + \chi_0 m')\, e^{\phi_0} \end{array} \right),
\end{equation}
Note that in the process of obtaining (2.19) and (2.20), the SO(2) angle
got fixed as,
\begin{eqnarray}
e^{i\alpha} &=& \left[(m - \chi_0 n) e^{\phi_0/2} + i n e^{-\phi_0/2}\right]
\tilde{\Delta}^{-1/2}_{(m,n)}\nonumber\\
&=& \left[(n' + \chi_0 m') e^{\phi_0/2} - i m' e^{-\phi_0/2}\right]
\bar{\Delta}^{-1/2}_{(m',n')}
\end{eqnarray}
{}From the above equation we find that the $\Delta$-factors associated with
the strings and the 5-branes are given as,
\begin{eqnarray}
\tilde{\Delta}_{(m, n)} &=& e^{\phi_0} (m - \chi_0 n)^2 + e^{-\phi_0} 
n^2\nonumber\\
\bar{\Delta}_{(m', n')} &=& e^{\phi_0} (n' + \chi_0 m')^2 + e^{-\phi_0} 
m'^2
\end{eqnarray}
Since the strings and 5-branes
are transformed simultaneously by the same SL(2, R) matrix, so 
it is clear from 
Eqs.(2.19) and (2.20) that the corresponding charges must be related as,
\begin{eqnarray}
(m, n) &=& k(a, b)\nonumber\\
(m', n') &=& k'(-b, a)
\end{eqnarray}
Here $(a,b)$ and $(k,k')$ are two pairs of relatively prime integers
which can be seen either from the general tension expression described
later in Eq.(2.31) or when we consider the special case bound states.
We, therefore, note that for the general ((F, D1), (NS5, D5)) configuration
i.e. when none of the integers are zero, $(m,n)$ and $(m',n')$ are not
relatively prime in contrast with the case when we consider either the
SL(2, Z) strings \cite{schone} or the SL(2, Z) 5-branes [9].
It can be easily checked that $\tilde{\Delta}_{(m,n)} = k^2 
\tilde{\Delta}_{(a,b)}$ and $\bar{\Delta}_{(m',n')} = k'^2 
\bar{\Delta}_{(-b,a)} = k'^2 \tilde{\Delta}_{(a,b)}$ are 
SL(2, Z) invariant.
Now once we find the SL(2, R)
transformation matrix $\Lambda_0$ given either by Eq.(2.19) or by (2.20),
we can obtain the general ((F, D1), (NS5, D5)) configuration by applying
the SL(2, R) transformation given in Eq.(2.10) on the initial (F, D5)
configuration. As mentioned earlier, the initial (F, D5) configuration
as given in Eqs.(2.1)--(2.6) should be modified by the appropriate $\Delta$
factors or more precisely, $q$ should be replaced by 
$\tilde{\Delta}^{1/2}_{(m,n)}$ and $s$ by $\bar{\Delta}^{1/2}_{(m',n')}$.
Keeping this in mind the final ((F, D1), (NS5, D5)) non-threshold bound
state is given by the following metric
\begin{eqnarray}
d s^2 =&& H'^{1/2} \, H^{1/4} \,
\left[ H^{-1} \left(- (d x^0)^2 +
(d x^1)^2\right)
+ H'^{-1} \left( (d x^2)^2 + (d x^3)^2 
+ (d x^4)^2 + (d x^5)^2\right)\right.\nonumber\\
&&\qquad\qquad \left. + d y^i d y^i\right],
\end{eqnarray}
with $i = 1, 2, 3, 4$; the dilaton,
\begin{equation}
e^{\phi} = e^{\phi_0} H^{- 1/2} H'',
\end{equation}
the axion,
\begin{equation}
\chi = \frac{\chi_0 + (H - 1) ab e^{-\phi_0} / \tilde{\Delta}_{(a,b)}}{H''},
\end{equation} 
and the rest of the non-vanishing fields,
\begin{eqnarray}
H_3^{(1)} &=&  - \frac{k}{\sqrt{k^2 + k'^2}}\tilde{\Delta}_{(a,b)}^{-1/2}\,  
e^{\phi_0}(a-\chi_0 b)\,d H^{-1}\wedge d x^0 
\wedge d x^1 - \frac{k' b \sqrt{2}\kappa_0 Q_0^5}{\Omega_3} \epsilon_3, 
\nonumber\\     
H_3^{(2)} &=&  \frac{k}{\sqrt{k^2 + k'^2}}\tilde{\Delta}_{(a,b)}^{-1/2}\,  
\left[e^{\phi_0}\chi_0(a-\chi_0 b) - e^{-\phi_0} b\right]
\,d H^{-1}\wedge d x^0 
\wedge d x^1 + \frac{k' a \sqrt{2}\kappa_0 Q_0^5}{\Omega_3} 
\epsilon_3,\nonumber\\ 
H_5 &=&  \frac{k k'}{k^2 + k'^2}   \, H'^{-2} 
\,d H \wedge d x^2 \wedge d x^3 \wedge d x^4 \wedge d x^5.
\end{eqnarray}
In the above the harmonic function $H = 1 + \frac{Q_5}{r^2}$, where
\begin{equation}
Q_5 =\sqrt{k^2 + k'^2} \tilde{\Delta}^{1/2}_{(a,b)} \frac{\sqrt{2}\kappa_0
Q_0^5}{2\Omega_3}
\end{equation}
$H'$ is another harmonic function where
\begin{equation}
H' = 1 + \frac{k'^2 Q_5/(k^2 + k'^2)}{r^2}
\end{equation}
and we have introduced a new harmonic function
\begin{equation}
H'' = 1 + \frac{b^2 e^{-\phi_0} Q_5 /\tilde{\Delta}_{(a,b)}}{r^2}
\end{equation}
 We note that the metric in (2.24) 
retains its form after SL(2, R) transformation except for the
introduction of the appropriate $\Delta$-factors as expected. The 5-form
field strength $H_5 = \ast H_5$ in (2.27) is SL(2, R) invariant. Also, from
(2.25) and (2.26) we find that as $r \rightarrow \infty$, $e^{\phi}
\rightarrow e^{\phi_0}$ and $\chi \rightarrow \chi_0$, the corresponding
asymptotic values as it should be. Let us now discuss how 
the (F, D5), (D1, NS5) as well as the degenerate cases (NS5, D5)
and (F, D1) bound
states can be obtained from this general configuration as special cases.

{}From the above solution given by (2.24)--(2.30),
we can obtain (F, D5) bound state by setting $a=1$ and $b=0$. Note that the
charges associated with the F-stirngs and the D5-branes are $k$, $k'$
respectively. As shown in [2], (F, D5) will form non-threshold bound states
only when $k$ and $k'$ are relatively prime integers.
Similarly,  the other bound state (D1, NS5) can be
obtained by setting $a=0$ and $b=1$ (also $\chi_0 = 0$). Here the charges 
associated with the
D-strings and NS5-branes are $k$ and $-k'$ respectively. 
However, since string charges are related to the 5-brane
charges as given in Eq.(2.23), we can get neither (F, NS5) nor (D1, D5) 
bound states and this is 
consistent with the fact that these configurations break 1/4 of the spacetime 
supersymmetries as can be inferred from that of the bound state of (D0, D4)
discussed in \cite{doukps,papt}. It should be
pointed out that for both $k$, $k'$ non-zero, the existence of $(m,n)$
string bound states seem to imply the existence of $(m',n')$ 5-brane
bound states. This is nice since the existence of 5-brane bound state is
not easy to establish considering the unrenormalizability of six dimensional
SYM theory and our poor understanding of the solitonic 5-branes. However,
some interesting scenario for the existence of 5-brane bound states has
been suggested in ref.\cite{witt}. The degenerate (NS5, D5) and (F, D1)
non-threshold bound state configurations can also be obtained from
((F, D1), (NS5, D5)) configuration given in Eqs. (2.24) -- (2.30) by
simply setting $k=0$, $k'=1$ and $k=1$, $k'=0$ respectively. In the
former case, we get the SL(2, Z) multiplet of 5-branes [9] (NS5, D5)
with magnetic charges $(-b, a)$ and in the latter case, we get SL(2, Z)
strings [6] with electric charges $(a, b)$ having additional isometries
in $x^2$, $x^3$, $x^4$, $x^5$ directions. Here $(a, b)$ are arbitrary
co-prime integers as can be shown for the strings [15] and 5-branes [9]
to form non-threshold bound states.

The expression for the string-frame tension of the general SL(2, Z) 
 bound state ((F, D1), (NS5, D5)) can be obtained by calculating the
mass per unit 5-brane volume. We can do so by following the steps given in
\cite{lurtwo} and by generalizing the ADM mass formula given in \cite{lu}. 
For a complete string-frame tension, we need to restore the $\phi_{B0}$ and
$\chi_{B0}$ which are set to zero from the outset for our above configuration.
In particular, we need to set $\phi_{B0} = \phi_0$ so that the string-frame 
 metric approaches Minkowski one  asymptotically as discussed in \cite{lurtwo}.
 With all these considerations,  the complete string-frame tension for this
 bound state takes the form:
\begin{equation}
T_5(k,k';a,b) = \frac{T_0^5}{g}\sqrt{\left[(k - \chi_{B0}\, k')^2 
g^2 + k'^2\right]
\left[(a - \chi_0 \,b)^2 + b^2 g^{-2}\right]}
\end{equation}
where $T_0^p = 1/[(2\pi)^p \alpha'^{(p+1)/2}]$ is the $p$-brane tension
unit and $g = e^{\phi_0}$ is the string coupling constant. This 
expression also clearly indicates that both the pairs of integers
$(k, k')$ and $(a, b)$ would have to be relatively prime if the SL(2, Z)
invariant state ((F, D1), (NS5, D5)) has to form a non-threshold 
bound state. Now it can be easily checked that the above formula
correctly reproduces the tensions of (F, D5) and (D1, NS5) bound states 
for $a=1$, $b=0$ and $a=0$, $b=1$ respectively (also $\chi_0 = 0$). 
Similarly, the tensions
for (NS5, D5) and (F, D1)\footnote{For (F, D1) case we have to multiply
the expression by $(2\pi)^4 \alpha'^2$ of four dimensional area
perpendicular to the strings since there are infinite number of (F, D1)
strings in the general bound state [1],[2].} can be obtained for
$k=0$, $k'=1$ and $k=1$, $k'=0$ respectively (also $\chi_{B0} = 0$).

\section{The Descendants of ((F, D1), (NS5, D5)) Bound State\protect\\}
\label{sec:dpb}

The T-duality rules for various BPS solutions along both longitudinal
and transverse directions in type IIA/IIB theories can be described by the
following table (for KK monopole, the transverse 
direction is taken to be the nut direction),

  \begin{center}
 \begin{tabular}{|c|c|c|}  \hline
  & Parallel
 & Transverse\\
 \hline
Dp& D(p $-$ 1) & D(p + 1)\\
F& W  & F \\
W& F  & W \\
NS5& NS5 &KK \\
KK & KK & NS5 \\
\hline\end{tabular}
\end{center}

In this table W, F, NS5 and KK  denote waves, 
fundamental strings,  NS fivebranes, and KK
monopoles, respectively, and they  are associated with 
NSNS fields.  Dp ($ - 1\le p \le 8$)\footnote{We do not consider D9 or
the space-time filling branes \cite{berg}.}
are the so-called D-branes and they are associated with  
RR fields.

Using the above table, we will give in this section, as a further example,  
the explicit space-time 
configuration of
((F, D2), (KK, D6)) bound state in type IIA theory  
by performing T-duality on 
the ((F, D1), (NS5, D5))
bound state along one of its transverse directions and then give the list
of other possible bound states towards the end. The general method 
of performing
T-duality has already been described at length in [2]. We here briefly
outline the method for completeness. T-duality along the transverse
direction is performed by the use of the so-called vertical dimensional
reduction and the diagonal or double dimensional oxidation method. Let us 
start with a $p$-brane (in our case it is the more complicated 
((F, D1), (NS5, D5)) brane) solution in type IIA (IIB) theory. We then use
the ``no-force'' condition of the BPS states to construct a multi-center
solution from the single center one with an infinite periodic array of
$p$-branes placed along the transverse direction. Then we take a continuum
limit to obtain the $p$-brane solution with one isometry along the would-be
compactified direction where the solution is now independent of this
coordinate. This process in turn reduces the dimensionality of the theory
to $D = 9$, known as the vertical dimensional reduction. Once we have
this solution, we perform the T-duality transformation on various fields
to write them from IIA (IIB) basis to IIB (IIA) basis. Then by the so-called
double-dimensional oxidation, we can simply
read off the $D = 10$ $(p+1)$-brane solution from the $D = 9$ solution. 
 One can also apply
T-duality along the longitudinal directions of the $p$-brane and obtain
new bound states by the method of diagonal reduction and vertical oxidation,
just opposite to the previous case.

{\bf ((F, D2), (KK, D6)) Bound State:} Here we assume that the (NS5, D5) 
in the original
((F, D1), (NS5, D5)) bound state is aligned along $x^1$, $x^2$, $x^3$, $x^4$
and $x^5$ direction and we apply T-duality along $x^6$ direction 
(assuming (F, D1) to
be aligned along the $x^1$-direction). Then 
following the procedure just outlined [2], we find that this bound state
in type IIA theory is given by the following Einstein metric,
\begin{eqnarray}
d s^2 =&& e^{\phi_0/8} H^{1/4} H'^{5/8} H''^{1/8}  
\left[ H^{-1} \left(- (d x^0)^2 +
(d x^1)^2\right) + H'^{-1}\left\{(dx^2)^2+\cdots +(dx^5)^2\right.\right.
\nonumber\\
&& \left.\left.  
 + e^{-\phi_0} H''^{-1} \left(d x^6 + k' b (\sqrt{2} 
\kappa_0 Q_0^6 / \Omega_2)
 (1 - \cos \theta) d \varphi\right)^2\right\}
 + d y^i d y^i\right],
 \label{eq:met}
\end{eqnarray}
where $\theta$ and $\varphi$ are the angular coordinates of 
$y^1, y^2$ and $y^3$ and 
$i = 1, 2, 3$; the dilaton,
\begin{equation}
e^{\phi} = e^{3\phi_0/4} H^{-1/2} H'^{- 1/4} H''^{3/4},
\end{equation}
and the rest of the non-vanishing fields,
\begin{eqnarray}
F_2 &=& \left(\chi_0 - \frac{a}{b}\right) dH''^{-1}\wedge dx^6 - k' a 
\frac{
\sqrt{2}\kappa_0 Q_0^6}{\Omega_2}  \epsilon_2\nonumber\\
F_3 &=&  - \frac{k}{\sqrt{k^2 + k'^2}} 
\tilde{\Delta}_{(a,b)}^{- 1/2} \,  
e^{\phi_0}(a-\chi_0 b)\,d H^{-1}\wedge d x^0 \wedge d x^1,\nonumber\\ 
F_4' &=& \frac{k k'}{\sqrt{k^2 + k'^2}} 
\tilde{\Delta}_{(a,b)}^{1/2} \frac{\sqrt{2}\kappa_0 Q_0^6}{\Omega_2} H^{-1}
d x^0 \wedge d x^1 \wedge \epsilon_2 \nonumber\\
&\,& + \frac{k}{\sqrt{k^2 + k'^2}} 
\tilde{\Delta}_{(a,b)}^{- 1/2} \, b e^{-\phi_0} \left( H H''\right)^{-1} 
d H \wedge d x^0 
\wedge d x^1 \wedge d x^6.  
\end{eqnarray}
Here the harmonic functions $H$, $H'$ and $H''$ are given as,
\begin{equation}
H = 1 + \frac{Q_6}{r}, \quad H' = 1 + \frac{k'^2 Q_6 /(k^2 + k'^2)}{r},
\quad {\rm and} \quad H'' = 1 + \frac{b^2 e^{-\phi_0} 
Q_6/\tilde{\Delta}_{(a,b)}}{r}
\end{equation}
where $Q_6 = \sqrt{k^2 + k'^2} \tilde{\Delta}^{1/2}_{(a,b)} 
\sqrt{2}\kappa_0 Q_0^6/\Omega_2$.

As discussed in the previous section we can obtain the bound states
(F, D6), (D2, KK) as well as the degenerate cases (F, D2), (KK, D6)
from this general bound state as special cases by simply setting
$a=1$, $b=0$\footnote{It can be easily checked from (2.26) that when
$b=0$, the first term of $F_2$ in Eq.(3.3) will not contribute.}; 
$a=0$, $b=1$; $k=1$, $k'=0$ and $k=0$, $k'=1$ respectively
in Eqs.(3.1) -- (3.4). Note that for the case of (F, D2) we have
additional isometries in $x^2$, $x^3$, $x^4$ and $x^5$ directions.
Again as before, we can not get the bound states (F, KK) and (D2, D6)
from the general bound state because of the charge relation Eq.(2.23).
This is consistent with the fact that these states preserve 1/4
spacetime supersymmetries.
Note that in order for the above metric 
Eq.\ (\ref{eq:met}) to be free from
conical singularity, $x^6$ should have a period of 
$4 \pi k' b (\sqrt{2} \kappa_0 Q_0^6 / \Omega_2)$.

A complete string-frame tension formula similar to Eq.(2.31) 
can also be written for the
general bound state ((F, D2), (KK, D6)) in the form:
\begin{equation}
T_6(k,k';a,b) = \frac{T_0^6}{g}\sqrt{\left[(k - \chi_{B0}\, k')^2 
g^2 + k'^2\right]
\left[(a - \chi_0 \, b)^2 + b^2 g^{-2}\right]}
\end{equation}
with $T_0^6$ as defined before. This expression reproduces the tensions
for the special case bound states (F, D6), (D2, KK), (F, D2), (KK, D6)
by setting $a=1$, $b=0$; $a=0$, $b=1$ (also $\chi_0 = 0$); $k=1$, $k'=0$ 
and $k=0$,
$k'=1$ (also $\chi_{B0} = 0$). As in the previous case, in order 
to get the correct tension
expression for (F, D2) we need to multiply the above expression by
the area $(2\pi)^4 \alpha'^2$.

At the level of supergravity solution as discussed in 
\cite{lurtwo}, we may expect
that we can make a further T-duality on ((F, D2), (KK, D6)) along one of the 
transverse directions of the D6 branes\footnote{If we T-dualize along the nut
direction, we are back to ((F, D1), (NS5, D5)).}. It is obvious from the
metric Eq.\ (\ref{eq:met}) that we have an isometry 
$\partial/\partial \varphi$.
But T-duality along this direction would result in a complicated metric 
which depends 
on the angle $\theta$\footnote{We would like to thank Chris Pope 
for pointing  
this out to us.}. We do not have a clear interpretation for the resulting
configuration. We therefore do not consider this T-duality here. Apart from
this possible T-duality, it is not obvious to us if we can have any other
simple T-duality as described above along a transverse direction.

Now we give the list of all possible descendants of ((F1, D1), (NS5, D5)) and 
((F, D2), (KK, D6)) by applying T-dualities
along various longitudinal directions of each of these two bound states.
We will follow the notation of ref.[8]. For example,
$({\rm T}_i: \mapsto)$ will denote T-duality along $i$-th direction. We assume
that the bound state ((F, D1), (NS5, D5)) is along $x^1$, $x^2$, $x^3$,
$x^4$ and $x^5$ directions and (F, D1)-strings are along $x^1$-direction.
Then according to the table given in the beginning of this section, 
we can  T-dualize each of the above ((F, Dp), (NS5/KK, D(p+4))), with
$1\leq p \leq 2$ (for $p$ = 1 we have NS5-state and for $p$ = 2 we
have KK-state), along longitudinal directions of the original (NS5, D5)-
branes to obtain new bound states. But since strings are along 
$x^1$-directions, we will get different bound states depending on whether
we T-dualize `1' direction first or not. T-dualizing along `1' we get, for
example, the following bound state
\begin{equation}
(({\rm F}, {\rm D1}), ({\rm NS5}, {\rm D5}))\,({\rm T}_1: \mapsto)\,
(({\rm W}, {\rm D0}), ({\rm NS5}, {\rm D4})).
\end{equation}
 We can also apply T-duality along longitudinal
directions other than `1' first and then apply T-duality along `1'. 
For example,
if we T-dualize along `5' first and then along `1' we get,
\begin{equation}
(({\rm F}, {\rm D1}), ({\rm NS5}, {\rm D5}))\,({\rm T}_5: \mapsto)\,
(({\rm F}, {\rm D2}), ({\rm NS5}, {\rm D4}))\,({\rm T}_1: \mapsto)\,
(({\rm W}, {\rm D1}), ({\rm NS5}, {\rm D3})).
\end{equation}
where the D2 and D4 in ((F, D2), (NS5, D4)) share only one common 
direction while
the D1 and D3 in ((W, D1), (NS5, D3) share no common directions. 
Repeating the above 
process to ((F, D2), (NS5, D4) along `4' first then along `1', we end up with
((F, D3), (NS5, D3)) and ((W, D1), (NS5, D3)). Continuing this, we 
have in general
((W, Dp), (NS5, D(4 $-$ p)) and ((F, D(p + 1)), (NS5, D(5 $-$ p)) for 
$0 \le p \le 4$.
Applying the similar process to ((F, D2), (KK, D6)), we have in general
((W, D(p + 1)), (KK, D(5 $-$ p)) and ((F, D(2 + p)), (KK, D(6 $-$ p)) 
for $0 \le p \le 4$.
 
This exhausts all the possibilities. In summary, 
we have the following list of all possible bound states which can be 
obtained by T-duality along one transverse and various longitudinal 
directions on
((F, D1), (NS5, D5)):
  
 \begin{center}
 \begin{tabular}{|c|c|}  \hline
 Bound States
 & no. common dir.\\
 \hline
((F, D(p + 1)), (NS5, D(5 $-$ p))) & 1\\
((F, D(p + 2)), (KK, D(6 $-$ p)))  &  2 \\
((W, Dp), (NS5, D(4 $-$ p)))  & 0 \\
((W, D(p + 1)), (KK, D(5 $-$ p))) & 1 \\
\hline\end{tabular}
\end{center}
  
Where the second column indicates the number of common directions shared by
the respective D-branes in the bound states. Also in the above $0\leq p\leq 4$.
If we write the above bound states in the form ((X, Y), (Z, V)), then from
the properties of ((F, D1), (NS5, D5)) discussed, we can get (X, V) by setting
$a = 1$, $b=0$, (Y, Z) by setting $a = 0$, $b=1$. The degenerate (Z, V) and
(X, Y) configurations can be obtained by setting $k=0$, $k'=1$ and
$k=1$, $k'=0$ respectively. 
These degenerate bound states have also been discussed, for example,
 in \cite{rust,cosp}. But we can not get the bound states (X, Z) and
(Y, V) because of the charge relations Eq.(2.23).

Now some of the states above belong to type IIB theory and so, we can apply
S-duality to those states to obtain new bound states. For example, from
((F, D3), (NS5, D3)) we can have (((F, D1), D3), ((NS5, D5), D3)) as a new
bound state. Similarly from other states also we can generate new bound
states by S-duality of type IIB theory. We can again apply T-duality on
these newly constructed bound states to obtain more bound states, then again
S-duality to those belonging to type IIB theory. Continuing this process
we can obtain all possible non-threshold bound states by S- and T-dualities
simply from the original (F, D1) strings. This process obviously will end
after a finite number of steps and thus we have finitely many bound states
in both type IIA and type IIB theories\footnote{Even if we count 
possible bound states
by applying T-dualities along the transverse directions of D6 
in ((F, D2), (KK,
D6)).}. Although at this stage we are unable
to count the exact number of bound states, but since there are finitely many
we believe that they may be related to the number of generators of the
largest finite U-duality group of type II theory i.e. E$_{8(+8)}$. We
speculate as in \cite{lurthree} that these bound states would 
form multiplets of E$_{8(+8)}$
U-duality symmetry in the yet unknown M- or U-theory. We will come back
to provide more evidence for this in the near future.

\section{Conclusion \protect\\}
\label{sec:c}

To summarize, by making use of SL(2, Z) symmetry of type IIB string theory
we have constructed in this paper a more general bound state of the type
((F, D1), (NS5, D5)) from the known (F, D5) configuration. There are infinite
number of $(m,n)$ strings forming bound state with $(m',n')$ 5-branes.
$(m,n)$ and
$(m',n')$ are respectively the integers corresponding to the charges
associated with the NSNS and RR gauge fields of the strings and 5-branes.
We have shown that a consistent quantization of charges of the strings
and 5-branes relates these integers as $(m,n) = k(a,b)$ and 
$(m',n') = k'(-b,a)$, where $(k,k')$ and $(a,b)$ are two pairs
of relatively prime integers. Thus the bound 
state ((F, D1), (NS5, D5)) is characterized by two pairs of integers 
$(k,k')$ and $(a,b)$. This seems to indicate that the
existence of string bound states  implies the existence of
5-brane bound states. From the explicit space-time configuration of
((F, D1), (NS5, D5)), we have shown how various bound states appear as
special cases. Thus we obtain (F, D5) and (D1, NS5) as well as the
degenerate cases (F, D1) and (NS5, D5)
bound states from here, but because of the charge relation between strings 
and 5-branes we can not get (F, NS5) and (D1, D5) bound states. This result
is consistent with the fact that (F, NS5) and (D1, D5) 
preserve 1/4 rather than
half of the spacetime supersymmetries. We have also given the tension
expression for the general ((F, D1), (NS5, D5)) non-threshold bound 
state which reduces to the correct expressions for the tensions of the
individual special case bound states by the proper choice of the integers
$(k,k';a,b)$.

The descendants of this bound state could be obtained by applying T-dualities
along various transverse and longitudinal directions as well as S-duality
of type IIB theory. We have given explicit space-time configuration of
((F, D2), (KK, D6)) in type IIA theory 
by applying T-duality in one of the transverse dierctions on 
((F, D1), (NS5, D5)). How the various bound states can be obtained as special
cases are also indicated. As in the previous case, we have given a
similar tension expression for this bound state. 
 Then we have given the list of all possible
bound states which can be obtained from ((F, D1), (NS5, D5)) and
((F, D2), (KK, D6)) by T-dualities. As we have mentioned, this is not the
end of the story. We can form new bound states by applying S-duality
to these T-dual bound states belonging to type IIB theory. T-duality 
can again be applied to these new set of bound states to generate another
new set. Then S-duality on those in type IIB theory will produce even more
bound states.
 Thus continuing this process we can generate all 
possible bound states, which will be finitely many, in both type IIA and
type IIB theories. All these states preserve one half of the space time
supersymmetries.
 We conjecture that these bound states would form 
multiplets of the largest finite U-duality group E$_{8(+8)}$ of yet unknown
M- or U-theory.
  
  \acknowledgments
  We would like to thank Chris Pope for discussions.
  JXL acknowledges the support of NSF Grant PHY-9722090. We would like
to thank the referee for raising an important issue which has helped us
to identify the more general charge relation Eq.(2.23) than has been
perceived by us in the earlier version of the paper.

  \end{document}